\documentclass{PoS}

\title{The Hottest Superfluid and Superconductor in the Universe: Discovery
and Nuclear Physics Implications}

\ShortTitle{The Hottest Superfluid and Superconductor in the Universe}

\author{\speaker{Wynn C. G. Ho},$^{a}$ Nils Andersson,$^a$
 Crist\'{o}bal M. Espinoza,$^b$ Kostas Glampedakis,$^c$
 Brynmor Haskell,$^{d,e}$ and Craig O. Heinke$^f$\\
        \llap{$^a$}School of Mathematics, University of Southampton,
Southampton, SO17~1BJ, United Kingdom\\
        \llap{$^b$}Jodrell Bank Centre for Astrophysics,
School of Physics and Astronomy, University of Manchester,
Manchester, M13~9PL, United Kingdom\\
        \llap{$^c$}Departamento de F\'{i}sica, Universidad de Murcia,
Murcia, E-30100, Spain\\
        \llap{$^d$}Astronomical Institute `Anton Pannekoek',
University of Amsterdam, Science Park~904, 1098~XH, Amsterdam, The~Netherlands\\
        \llap{$^e$}Max Planck Institute for Gravitational Physics
(Albert Einstein Institute), Am~M\"{u}hlenberg~1, D-14476, Golm, Germany\\
        \llap{$^f$}Department of Physics, CCIS~4-183, University of Alberta,
Edmonton, AB, T6G~2E1, Canada\\
        E-mail: \email{wynnho@slac.stanford.edu}}

\abstract{
We present recent work on using astronomical observations of neutron stars
to reveal unique insights into nuclear matter that cannot be obtained from
laboratories on Earth.
First, we discuss our measurement of the rapid cooling of the youngest neutron
star in the Galaxy; this provides the first direct evidence for superfluidity
and superconductivity in the supra-nuclear core of neutron stars.
We show that observations of thermonuclear X-ray bursts on neutron stars
can be used to constrain properties of neutron superfluidity and neutrino
emission.
We describe the implications of rapid neutron star rotation rates on aspects
of nuclear and superfluid physics.
Finally, we show that entrainment coupling between the neutron superfluid
and the nuclear lattice leads to a less mobile crust superfluid;
this result puts into question the conventional picture of pulsar glitches
as being solely due to the crust superfluid and suggests that the core
superfluid also participates.
}

\FullConference{Xth Quark Confinement and the Hadron Spectrum\\
                 8--12 October 2012\\
                 TUM Campus Garching, Munich, Germany}

\begin{document}

\section{Discovery of superfluid cooling of the Cassiopeia A neutron star}
\label{sec:casa}

Neutron stars are created in the collapse and supernova explosion of massive
stars, and they begin their lives very hot
(with $kT>10\mbox{ MeV}$) but cool rapidly through the emission of neutrinos.
This neutrino emission depends on uncertain physics at the supra-nuclear
densities ($\gtrsim 0.08\mbox{ fm$^{-3}$}$) of the neutron star core
\cite{tsuruta98,yakovlevpethick04,pageetal06,yakovlevetal11}.
Current theories indicate that the stellar core may contain exotica,
such as hyperons and deconfined quarks,
and matter may be in a superfluid/superconducting state
\cite{migdal59,haenseletal07,lattimer12}.
By observing the cooling of neutron stars and comparing their temperatures to
theoretical models, we can constrain the nuclear physics properties that
govern the stellar interior.

The compact object at the center of remnant of the Cassiopeia~A supernova
was discovered in \textit{Chandra X-ray Observatory} first-light observations
\cite{tananbaum99} and subsequently identified as a neutron star
\cite{hoheinke09}.
The supernova explosion is estimated to have occurred in the year $1681\pm 19$
\cite{fesenetal06};
this makes the Cassiopeia~A neutron star the youngest-known neutron star
at an age of $\approx 330\mbox{ yr}$.
A steady temperature decline of four~percent was found using \textit{Chandra}
observations taken during the last 10~years \cite{heinkeho10}.
If the rapid decline is due to passive neutrino cooling, then this is the
first direct evidence for superfluidity and superconductivity in the core
of a neutron star \cite{pageetal11,shterninetal11}.

\begin{figure}[htb]
\centering
\resizebox{!}{0.30\textheight}{\includegraphics{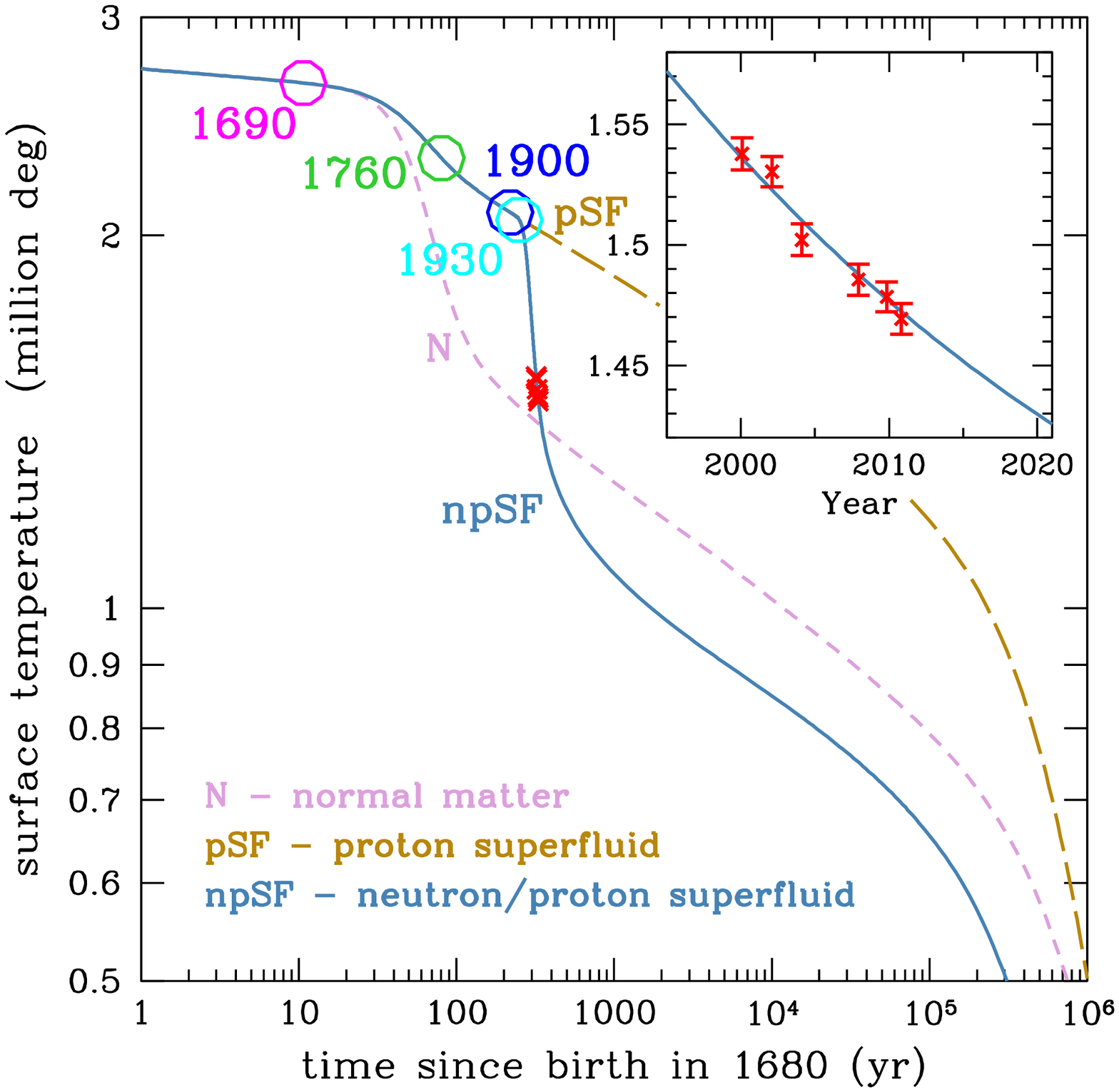}\hspace{3.0cm}\includegraphics{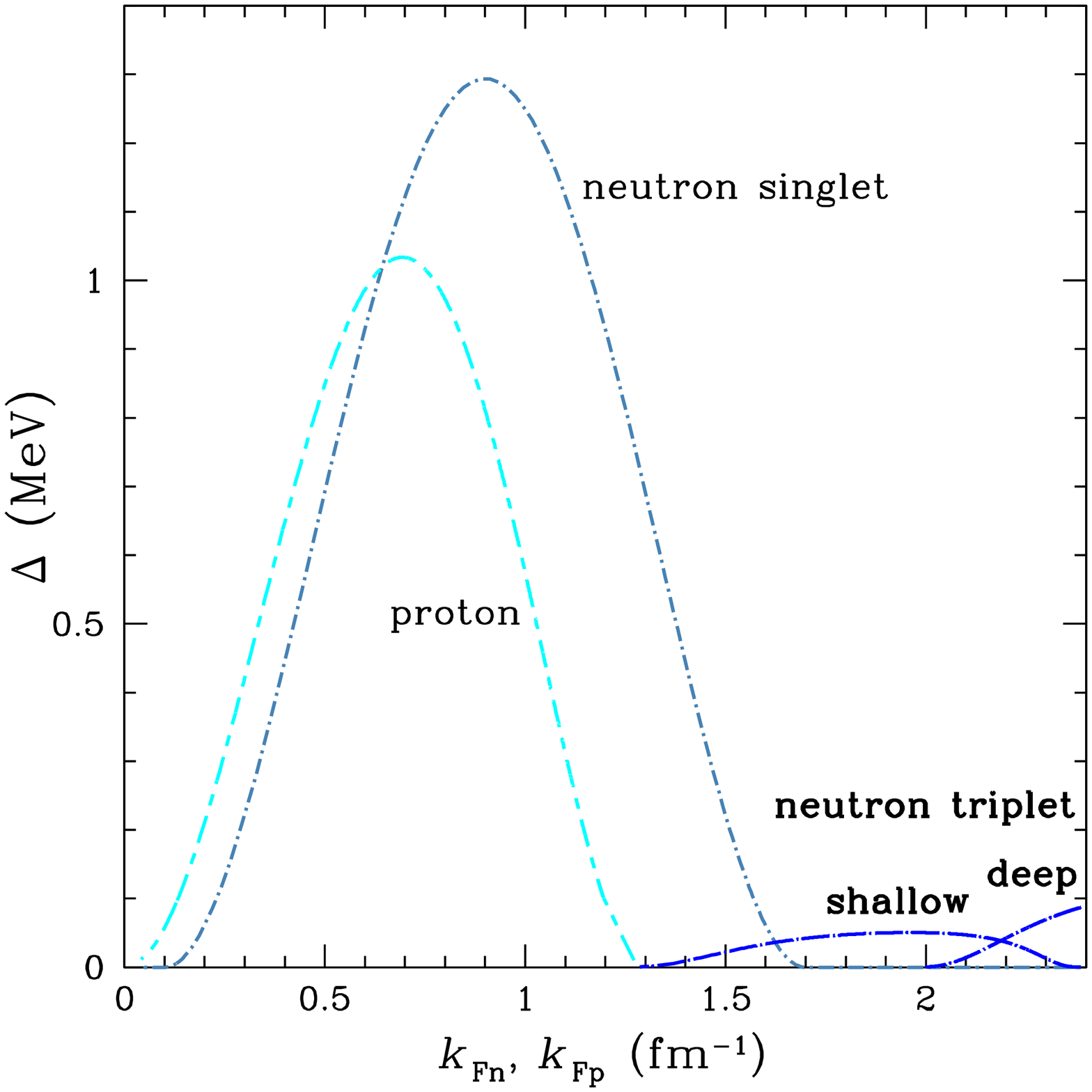}}
\caption{
Left: Theoretical models of neutron star cooling with
superfluid neutrons and protons (npSF $-$ solid),
normal neutrons and superfluid protons (pSF $-$ long-dashed), and
normal neutrons and protons (N $-$ short-dashed).
Circles indicate temperature for the npSF model at particular times/years.
Crosses are \textit{Chandra X-ray Observatory} measurements of the
Cassiopeia~A neutron star.
Right:
Superfluid pairing gap energies as a function of Fermi wavenumber for
neutrons $k_\mathrm{Fn}$ and protons $k_\mathrm{Fp}$.
The maximum neutron triplet gap is taken to be \textit{either} the
shallow model or the deep model, and each triplet gap leads to cooling that
can fit the Cassiopeia~A data (see \cite{pageetal11} and \cite{shterninetal11},
respectively).
}
\label{fig:casa}
\end{figure}

The left panel of Fig.~\ref{fig:casa} (in particular, see inset) shows
\textit{Chandra} temperature measurements of the Cassiopeia~A neutron star
from 1999 to 2010 \cite{heinkeho10,shterninetal11}.
Figure~\ref{fig:casa} also shows surface temperatures for three theoretical
models of neutron star cooling:
``N $-$ normal matter'' corresponds to neutron star matter that does not
contain any sort of superfluid,
``pSF $-$ proton superfluid'' is for superfluid protons in the core,
and ``npSF $-$ neutron/proton superfluid'' is for superfluid neutrons and
protons in the core.
Note the difference between the cooling behavior of models with
normal matter (N) and matter containing superfluids (pSF or npSF)
after $\approx 40\mbox{ yr}$.
In the latter models, a proton superconductor forms soon after neutron star
formation, and this suppresses neutrino emission,
so that the cooling rate is weaker than for normal matter.
This enables the star to stay relatively warm, leading to a rapid
temperature drop once neutrons become superfluid
\cite{gusakovetal04,pageetal04}.
The model with superfluid neutrons and protons (npSF) fits the data
at an age of a few hundred years.
The four circles trace the cooling curve predicted by this model from about
10 years after the supernova explosion (SN in $\sim 1680$) to about the time
when neutrons become superfluid in the core:
(1) At early ages, the neutron star core cools so rapidly by neutrino emission
that the crust does not have time to react. Thus the crust is hotter than the
core in 1690 (age $\approx 10\mbox{ yr}$; protons are superconducting by this
time), and the surface temperature declines very slowly.
(2) The surface temperature eventually reacts to the ``cooling wave''
that sweeps through the crust and starts to drop off more quickly.
After 1760, the temperature becomes almost constant throughout the star.
(3) Then in $\approx$ 1900, the interior temperature drops below the critical
value for a neutron superfluid to form and enhanced neutrino emission occurs
in the core, as neutron Cooper pairs form.
Energy is lost as the neutrinos are emitted, causing the core to cool off
and another cooling wave to travel outwards. As neutrons in large regions of
the core become superfluid, the surface temperature drops off quickly,
beginning in $\approx 1930$ (i.e., start of ``Great Depression'') and
continuing through the present date.
See Fig.~2 of \cite{hoetal12}, which shows evolution of interior temperature
$T(\rho)$ and transition to neutron superfluidity.

Monitoring of the temperature decline will allow improved constraints on the
(1) critical temperature for neutron triplet pairing $T_\mathrm{nt}$
(maximum pairing gap energy; see Fig.~\ref{fig:casa} and Fig.~1 of
\cite{shterninetal11}),
(2) suppression due to collective effects of the axial vector current
for pair formation (see Fig.~3 of \cite{shterninetal11}),
and (3) neutron star mass and nuclear equation of state
(see Fig.~1 of \cite{shterninetal11} and Fig.~4 of \cite{pageetal11}).

Guided by the discovery of a superfluid and superconductor in the
Cassiopeia~A neutron star,
we examine three examples where measurement of the pairing gap energies
has possible effects or where further constraints may be obtained.

\section{Nuclear X-ray bursts and neutrino cooling by neutron superfluid}
\label{sec:bursts}

In contrast to the neutron star in Cassiopeia~A, many old neutron stars are
found in binary systems.
These binaries can be seen in X-rays, which are produced when material from
the companion star accretes onto the neutron star.
If the companion has a low mass, the systems are known as low-mass X-ray
binaries (LMXBs).
Many LMXBs undergo bright X-ray bursts due to unstable thermonuclear burning
of hydrogen and/or helium in the surface layers of the neutron star
\cite{lewinetal93,gallowayetal08}.
Bursts are sometimes observed to recur in individual sources,
and recurrence times between multiple bursts span a wide range,
from minutes to days \cite{gallowayetal08,keeketal10}.
However, recurrence times $\lesssim 1\mbox{ hr}$ are too short for the
neutron star to accrete enough fuel for subsequent bursts
\cite{lewinetal93,woosleyetal04}.

\begin{figure}[htb]
\centering
\resizebox{!}{0.29\textheight}{\includegraphics{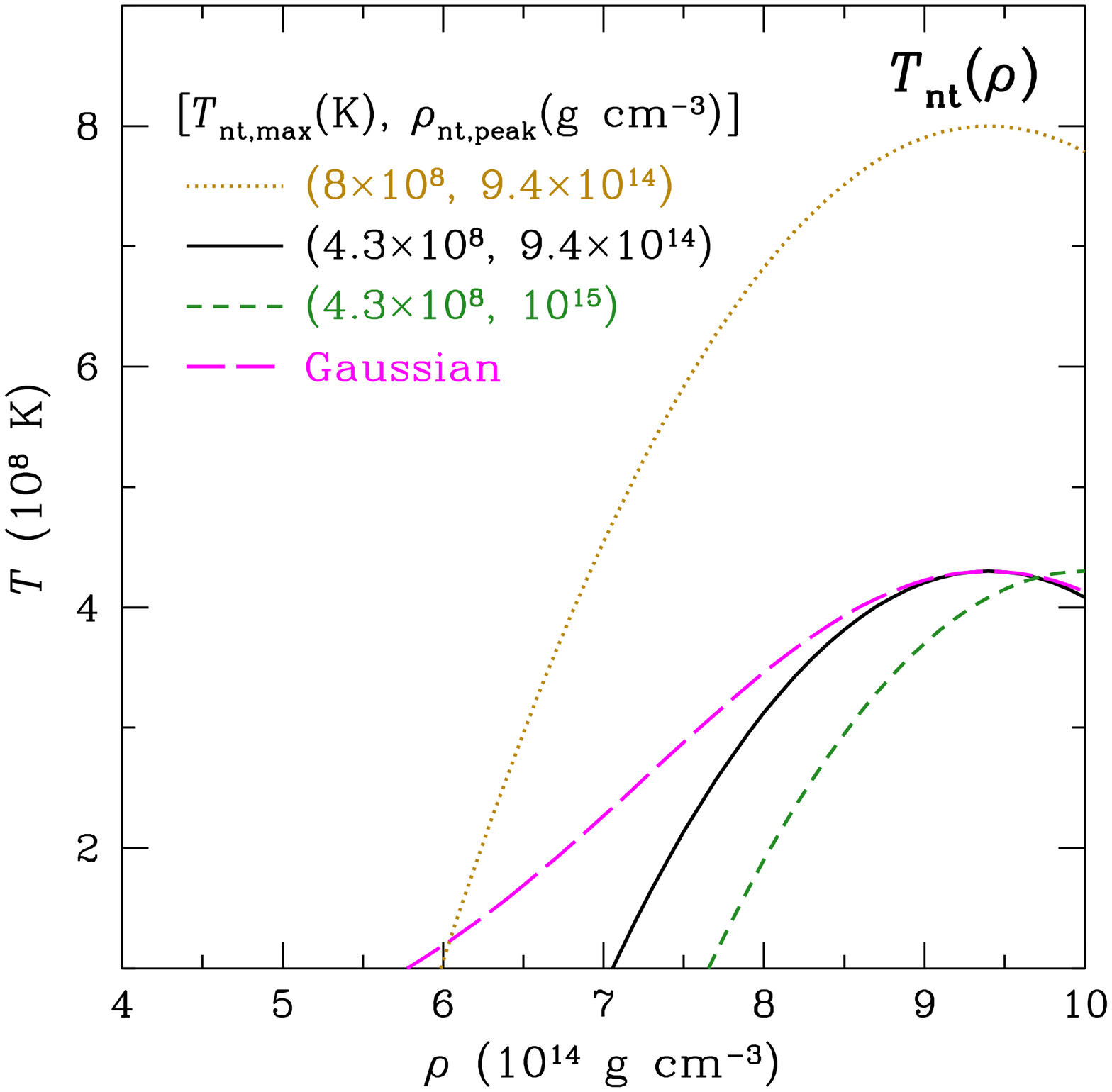}\hspace{3.0cm}\includegraphics{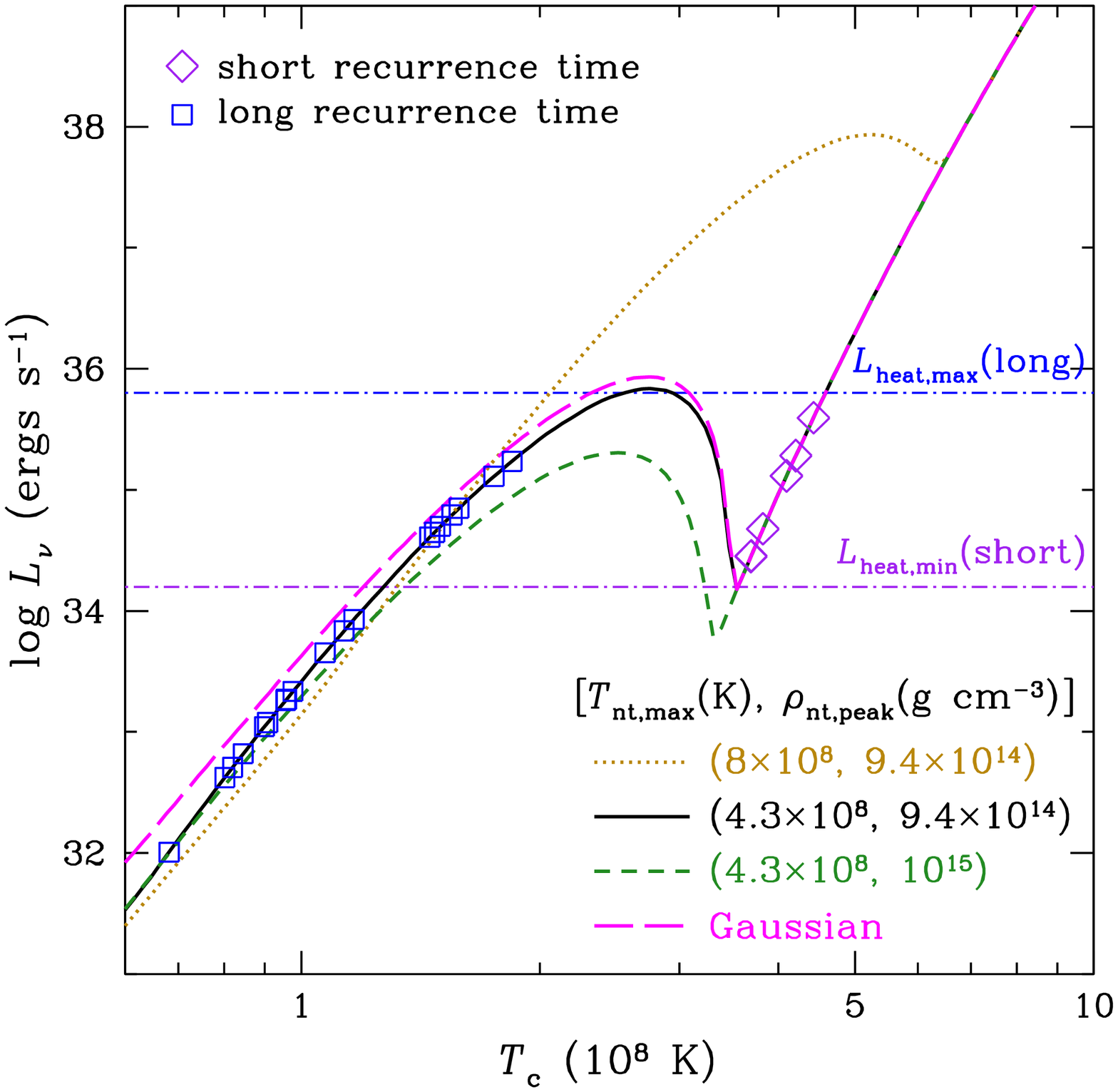}}
\caption{
Left: Simple models of the neutron triplet critical temperature
$T_\mathrm{nt}(\rho)$.
Right:
Neutrino luminosity as a function of neutron star core temperature,
where the different curves are $L_\nu$ calculated using the models of
$T_\mathrm{nt}(\rho)$ shown in the left panel.
Upper (lower) horizontal dot-dashed line is the highest (lowest)
observed $L_\mathrm{heat}$ from among all long (short) recurrence time bursts.
Squares and diamonds are where
$L_\mathrm{heat}=L_\nu$
(with $4.3\times 10^8\mbox{ K},9.4\times 10^{14} \mbox{ g cm$^{-3}$})$
for each LMXB with long and short recurrence times, respectively.
}
\label{fig:bursts}
\end{figure}

In \cite{ho11}, we revisit the method used to infer core temperatures
$T_\mathrm{c}$ of neutron stars in LMXBs.
Compression by accreted matter induces nuclear reactions in the deep crust,
which release $\approx 1.5$ or $1.9\mbox{ MeV nucleon$^{-1}$}$
\cite{haenselzdunik08}, and this heats the core directly
by a luminosity $L_\mathrm{heat}\approx 0.0078\,L_\mathrm{acc}$,
where $L_\mathrm{acc}$ is the time-averaged X-ray luminosity of the LMXB
\cite{brownetal98,brown00}.
Figure~\ref{fig:bursts} shows the measured heating rate $L_\mathrm{heat}$,
as well as the theoretical neutrino luminosity $L_\nu$, which depends on
the neutron triplet pairing gap energy (or critical temperature
$T_\mathrm{nt}$).
The intersection of the curves $L_\mathrm{heat}$ and $L_\nu$ yields the
neutron star core temperature.
We see that the core temperature in relatively high luminosity LMXBs may
not be uniquely determined.
If $T_\mathrm{nt,max}\lesssim 8\times 10^8\mbox{ K}$, there can be two
thermally stable values\footnote{There can be three values of $T_\mathrm{c}$
that intersect each horizontal $L_\mathrm{heat}$.  But the intermediate
temperature is thermally unstable since a temperature decrease leads
to an increase in neutrino luminosity which causes even more rapid cooling.}
of the core temperature associated with a single observed
accretion luminosity.
For example, there is a factor of $< 3$ difference in the inferred
$T_\mathrm{c}$ if
$L_\mathrm{acc}\sim (0.2-9)\times 10^{37}\mbox{ ergs s$^{-1}$}$ and
$T_\mathrm{nt,max}=4.3\times 10^8$~K.
The luminosities of all LMXBs that show short burst recurrence time lie
within this range \cite{keeketal10}.
To highlight this point, we place the LMXBs with ``short time'' on the
high-temperature branch and LMXBs with ``long time'' on the low-temperature
branch.
Thus the sources with short time and higher temperatures have normal neutrons
in the stellar core, while sources with long time and lower temperatures have
superfluid neutrons.

If short burst recurrence time LMXBs do indeed possess hotter core
temperatures, then measurements of the minimum and maximum accretion
luminosities of bursts from short time LMXBs and long time LMXBs,
respectively, can be used to constrain the neutron superfluid critical
temperature $T_\mathrm{nt}(\rho)$.
This is illustrated in Fig.~\ref{fig:bursts}, where it is clear that the
accretion luminosities for LMXBs can constrain $T_\mathrm{nt,max}$ and
$\rho_\mathrm{nt,peak}$,
while the width of $T_\mathrm{nt}(\rho)$ is not as important in determining
the qualitative behavior of $L_\nu$.

\section{Neutron star spin and damping of r-mode oscillations}
\label{sec:rmode}

One of the main mechanisms that is expected to affect the spin evolution
of an accreting neutron star in a LMXB is the instability associated with
r-modes, which are a class of oscillations in a star whose restoring force
is the Coriolis force.
The emission of gravitational waves can excite r-modes in the stellar core
and cause the oscillations to grow \cite{andersson98}.
The r-mode instability is interesting for many reasons, mainly because the
associated gravitational wave signal may be detectable,
but also because its understanding requires knowledge from a wide range of
nuclear physics.
A primary agent that enters the r-mode discussion is damping mechanisms
related to shear and bulk viscosities and exotica like hyperons, quarks, and
superfluid vortices \cite{anderssonkokkotas01}
(see also C.J.~Horowitz, this volume).
The instability depends primarily on the neutron star spin rate
$\nu_\mathrm{s}$ and core temperature $T_\mathrm{c}$.
This leads to an instability ``window,'' determined by a critical curve in the
$\nu_\mathrm{s}$-$T_\mathrm{c}$ plane, inside which the instability is active.
What has not been appreciated is that this leaves the majority of the
observed systems significantly inside the instability window:
rapidly rotating neutron stars (i.e., pulsars) should not possess spin rates
at their observed levels \cite{hoetal11,haskelletal12}.

\begin{figure}[htb]
\centering
\includegraphics[width=\textwidth]{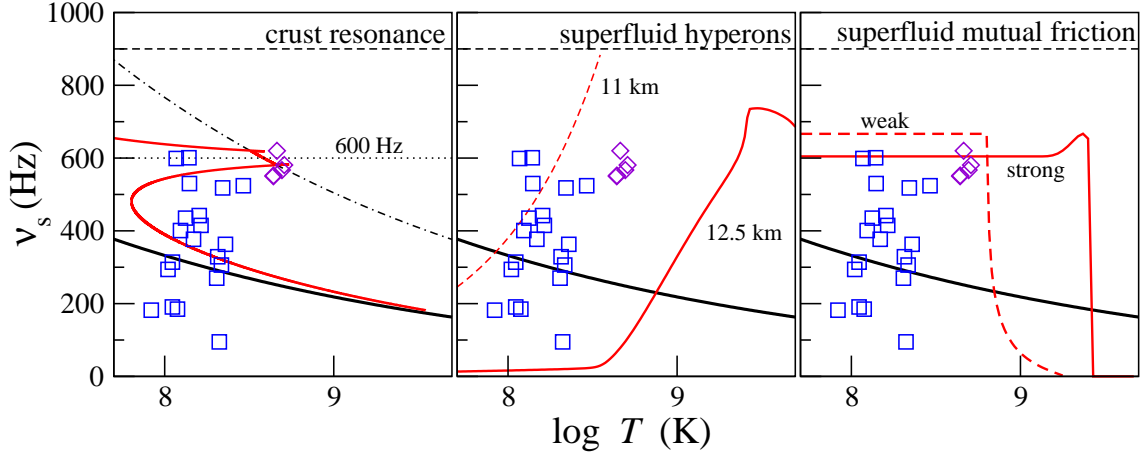}
\caption{
Three scenarios that could explain r-mode stability in the observed LMXBs
(squares and diamonds; see Fig.~2).
R-mode growth is stable below (i.e., at lower $\nu_\mathrm{s}$) the various
curves, while the dashed lines at 900~Hz indicate the break-up limit.
Left: Crust mode resonance at 600~Hz.
Middle: Superfluid hyperons (based on \cite{haskellandersson10} with
$\chi=0.1$).
Right: Strong vortex mutual friction (based on the strong/weak superfluidity
models from \cite{haskelletal09} with $\mathcal{B}\approx 0.01$).
}
\label{fig:rmode}
\end{figure}

One solution to this dilemma is to change the window so that r-mode growth
is stabilized at relatively high spin rates.
However in order to do this, a revision of our understanding of damping
mechanisms is required.
In \cite{hoetal11}, we explore the possibilities.
For example, there may be resonances between the r-mode and torsional
oscillations of the elastic crust \cite{levinushomirsky01}.
Such resonances could have a sizeable effect on the instability window.
Figure~\ref{fig:rmode} shows an example;
the illustrated instability window has a broad resonance at 600~Hz, which
is the typical frequency of the first overtone of pure crustal modes.

Another possibility is an instability spin frequency that increases with
temperature in the range of interest here \cite{anderssonetal02,wagoner02}.
If this is the case, then neutron stars may evolve to a quasi-equilibrium
where the r-mode instability is balanced (on average) by accretion and
r-mode heating is balanced by cooling.
This solution is interesting because it predicts persistent 
(low-level) gravitational radiation.
Figure~\ref{fig:rmode} shows a model using hyperon bulk viscosity
suppressed by superfluidity.
This explanation has a major problem though.
It must be able to explain how observed pulsars with millisecond spin rate
emerge from accreting systems.
Once the accretion phase ends, the neutron star will cool, enter the
instability window, and spin down to $\sim 300$~Hz.
In other words, it would be very difficult to explain the formation of a
pulsar spinning at 716~Hz \cite{hesselsetal06}.

An intriguing possibility involves mutual friction due to vortices
in a rotating superfluid. The standard mechanism (electrons scattered off
of magnetized vortices) is too weak to affect the instability window
\cite{lindblommendell00}.
However, if we increase (arbitrarily) the strength of this mechanism by
a factor $\sim 25$, then mutual friction dominates the damping, as shown
in Fig.~\ref{fig:rmode}.
Moreover this would set a spin threshold for instability similar to the highest
observed $\nu_\mathrm{s}$ and would allow systems to remain rapidly rotating
after accretion shuts off.  Enhanced friction may result from the
interaction between vortices and proton fluxtubes in the outer core.

\section{Pulsar glitches and neutron superfluidity in the crust}
\label{sec:glitch}

Mature neutron stars tend to have extremely stable spin rates, with some
pulsars possessing a timing stability that rivals the best terrestrial
atomic clocks.
However, young neutron stars may behave in a less ordered fashion.
In particular, many young pulsars exhibit regular glitches, where the
observed spin rate suddenly increases \cite{espinozaetal11}.
The consensus view is that these events are due to a superfluid component
in the stellar interior \cite{baymetal69}.
Anderson \& Itoh \cite{andersonitoh75} envisaged a glitch as a tug-of-war
between the tendency of the neutron superfluid to match the spindown rate
of the rest of the star by expelling vortices and the impediment experienced
by moving vortices that are pinned to crust nuclei.
Strong vortex pinning prevents the superfluid from spinning down
and creates a spin lag with respect to the rest of the star
(which slows by magnetic dipole radiation).
This situation cannot persist forever.
The increasing spin lag leads to a build-up in the Magnus force exerted on
the vortices.
Above a threshold, pinning can no longer be sustained, vortices break free,
and excess angular momentum is transferred to the crust.
This leads to the observed spin-up, i.e., glitch.

\begin{figure}[htb]
\centering
\resizebox{!}{0.29\textheight}{\includegraphics{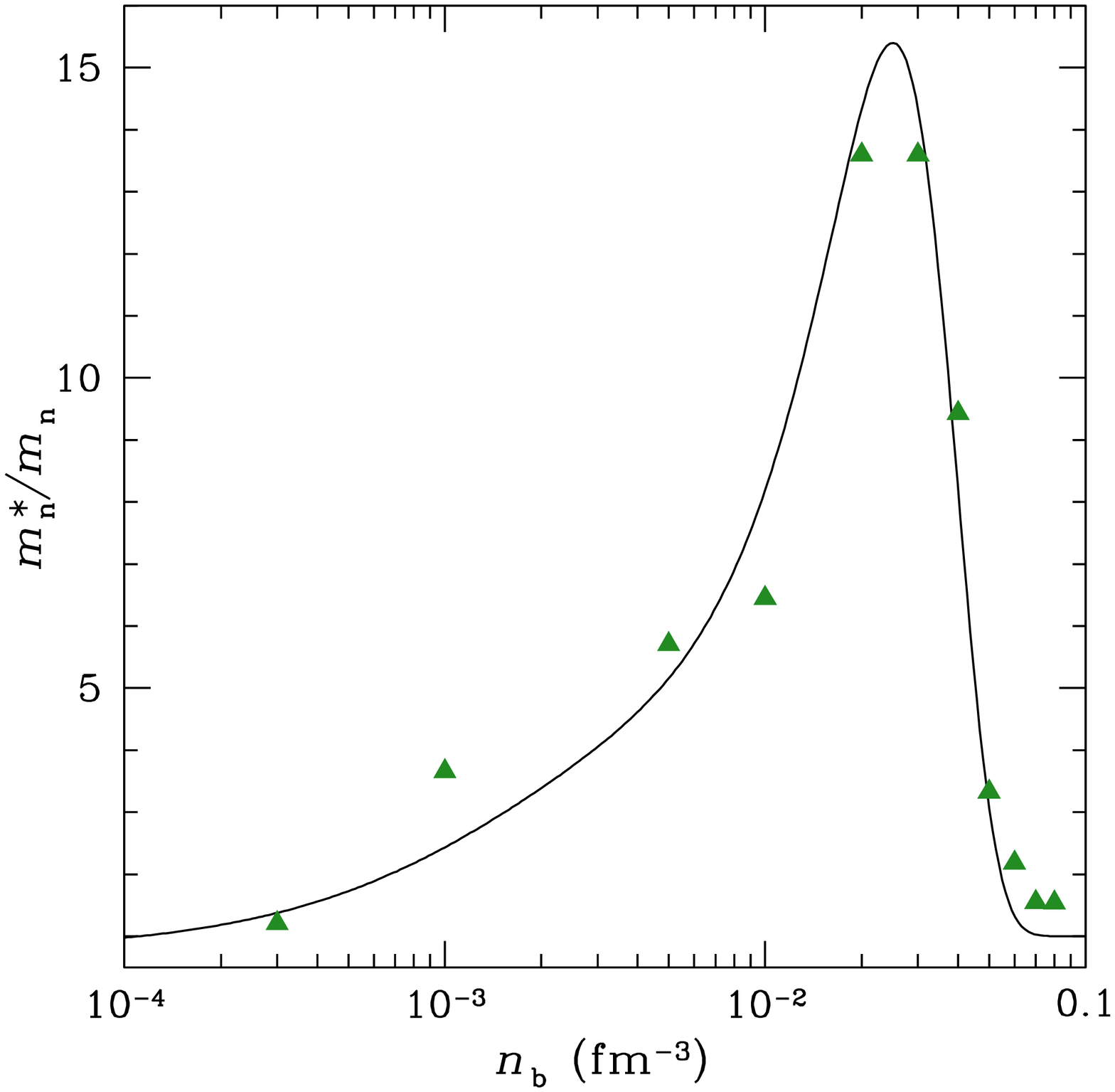}\hspace{3.0cm}\includegraphics{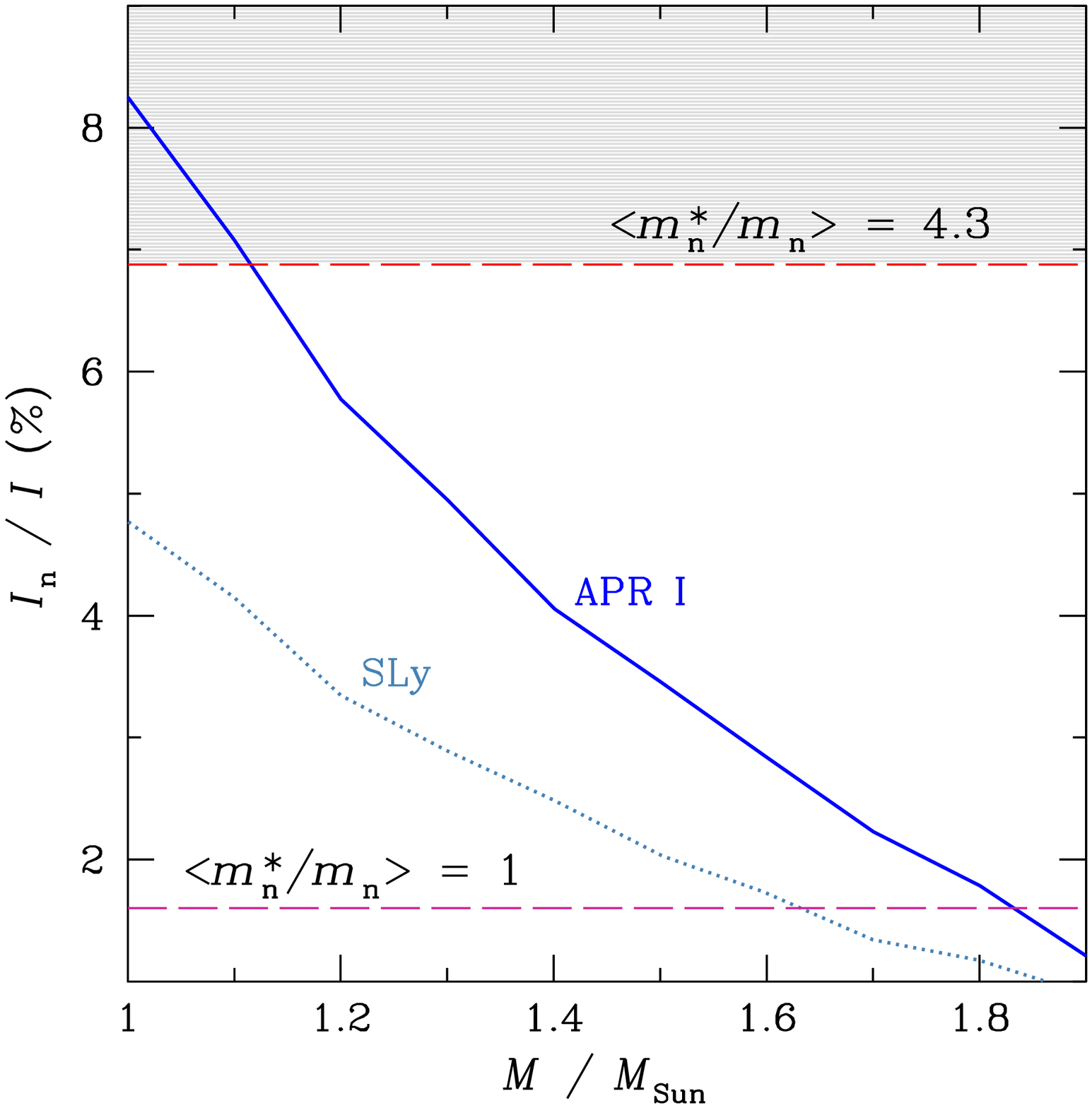}}
\caption{
Left: Neutron effective mass as a function of baryon number density.
Triangles are from \cite{chamel12}, while the curve is a fit from
\cite{anderssonetal09}.
Right: Moment of inertia ratio $I_\mathrm{n}/I$ as a function of mass
for neutron star models built from the APR~I (solid) and SLy (dotted)
nuclear equations of state.
If glitches in the Vela pulsar are to be explained solely by a crust
superfluid, then the moment of inertia ratio must satisfy
$I_\mathrm{n}/I\gtrsim 0.016\times(<m_\mathrm{n}^\ast>/m_\mathrm{n})\approx
0.07$, where the average effective mass is calculated from that shown in
the left panel;
also shown is the constraint when entrainment is not taken into account,
i.e., when $<m_\mathrm{n}^\ast>/m_\mathrm{n}=1$.
}
\label{fig:glitch}
\end{figure}

A previous analysis by \cite{linketal99} suggests that glitches
involve a superfluid reservoir with moment of inertia
$I_\mathrm{n}/I\sim 1\%$, where $I$ and $I_\mathrm{n}$ are the moments of
inertia of the entire star and the neutron superfluid component, respectively.
The similarity of the inferred $I_\mathrm{n}$ to the
theoretically estimated moment of inertia of the crust (which is dominated by
free neutrons in the inner crust) for realistic nuclear equations of state
\cite{ravenhallpethick94} supports the idea that glitches involve only the
crust region.
In \cite{anderssonetal12}, we show that this logic breaks down when one
accounts for non-dissipative entrainment coupling between the neutron
superfluid and the crust lattice, an effect which can be expressed in terms
of an effective neutron mass $m_\mathrm{n}^\ast$.
Recent work indicates that this effective mass may be significantly larger
than the bare neutron mass $m_\mathrm{n}$ \cite{chamel12}
(see Fig.~\ref{fig:glitch}).
This implies a decreased superfluid mobility with respect to the
lattice and the need for a larger angular momentum reservoir for glitches. 
Combining the latest glitch data \cite{espinozaetal11} with
a general relativistic multifluid model that includes entrainment, we find
that the requisite superfluid moment of inertia is above the capacity of the
crust superfluid \cite{anderssonetal12} (see Fig.~\ref{fig:glitch}).
Some solutions are briefly discussed below (see also \cite{anderssonetal12}).

One possible explanation could be that the superfluid in the core is involved
in the glitch (see, e.g., \cite{haskelletal12b}),
and the combined superfluid moment of inertia reservoir is just
large enough to explain the observations.
If this fine-tuning resolves the problem, then a more detailed calculation
would constrain the singlet pairing gap for neutrons.
This would be an interesting complement to the constraints on core
superfluids (singlet protons and triplet neutrons) discussed in
Sections~\ref{sec:casa}--\ref{sec:rmode}.

Another solution could be the result of superfluid behavior at the
crust-core transition.
Unless the superfluid is confined to the crust, one would have to explain
why the crust component decouples from the core during the glitch event.
This would be particularly vexing if the singlet pairing gap is such that
the neutron superfluid reaches far into the core.
A central issue concerns the nature of superfluid vortices extending across
this interface. The standard picture is that vortices are magnetized in the
core \cite{alparetal84}, due to entrainment and the presence of
superconducting protons, but not in the crust.
This suggests a more complicated transition behavior than is usually assumed.

\vspace{1em}
Full details of the work presented in Sections~\ref{sec:casa}--\ref{sec:glitch}
can be found in \cite{shterninetal11} (see also \cite{pageetal11}), \cite{ho11},
\cite{hoetal11} (see also \cite{haskelletal12}), and \cite{anderssonetal12}
(see also \cite{chamel13}), respectively.

\vspace{1em}
WCGH is indebted to Daniel~Patnaude, Peter~Shternin, and Dmitry~Yakovlev for
assistance.

\end{document}